\documentclass[a4paper,11pt]{article}
\pdfoutput=1 

\usepackage{jinstpub} 
\usepackage{comment}
\usepackage{lineno}
\usepackage{graphicx}
\usepackage{subcaption}
\usepackage{comment}

\title{\boldmath Graph Neural Networks for reconstruction and classification in KM3NeT}

\author[a,1]{S.Reck,\note{Corresponding author.}}
\author[b]{D. Guderian,}
\author[c, e]{G. Vermari\"en,}
\author[d, e]{A. Domi}

\affiliation[a]{Universität Erlangen-Nürnberg, Erwin-Rommel-Str. 1, 91058 Erlangen, Germany\\Erlangen Centre for Astroparticle Physics (ECAP)}
\affiliation[b]{WWU Münster,Wilhelm-Klemm-Straße 9, 48149 Münster, Germany}
\affiliation[c]{Leiden Observatory, Leiden University,Niels Bohrweg 2, Leiden, The Netherlands}
\affiliation[d]{University of Amsterdam, Institute of Physics,
Science Park 904, Amsterdam, The Netherlands}
\affiliation[e]{Dutch National Institute for Subatomic Physics (Nikhef),
Science Park 105, Amsterdam, The Netherlands}

\emailAdd{stefan.reck@fau.de}

\abstract{KM3NeT, a neutrino telescope currently under construction in the Mediterranean Sea, consists of a network of large-volume Cherenkov detectors. 
Its two different sites, ORCA and ARCA, are optimised for few GeV and TeV-PeV neutrino energies, respectively. 
This allows for studying a wide range of physics topics spanning from the determination of the neutrino mass hierarchy to the detection of neutrinos from astrophysical sources. 

Deep Learning techniques provide promising methods to analyse
the signatures induced by charged particles traversing the detector.
This document will cover a Deep Learning based approach
using Graph Convolutional Networks to classify and reconstruct events in both the ORCA and ARCA detector.
Performance studies on simulations as well as applications to real data will be presented, together
with comparisons to classical approaches.
}


\collaboration[c]{on behalf of the KM3NeT collaboration}

\proceeding{VLVNT 2021 - Very Large Volume Neutrino Telescope Workshop\\
  18-21 May 2021\\
  Valencia}

\begin{document}
\maketitle
\flushbottom

\section{Introduction}
\label{sec:intro}

KM3NeT is a network of neutrino telescopes under construction in the Mediterranean Sea \cite{Adrian-Martinez2016}, consisting of two detectors: ORCA, optimised for oscillation studies with atmospheric neutrinos in the 1 to 100 GeV range, and ARCA, optimised for cosmic neutrino searches in the TeV to PeV energy range. 
Charged particles produced in neutrino interactions with water emit Cherenkov light, which is detected by 3" photomultiplier tubes (PMTs) hosted in high-pressure glass spheres (DOMs). 31 PMTs are stored in a single DOM, and 18 DOMs are fixed to a long vertical detection unit (DU) that is anchored to the sea floor. Each of the three building blocks of KM3NeT will consist out of 115 DUs.
The detected light on each PMT (hit) can be used for reconstructing the particles' properties, such as their energy or direction. A hit holds information about the time and the xyz-position of the PMT that recorded the light, as well as its pointing direction. Together, the hits of an event are the input to reconstruction and classification algorithms. Since the data recorded by KM3NeT closely resembles point clouds, Graph Neural networks (GNNs) are a natural choice for the architecture.

In the input to the GNN, the information of each single hit becomes the node feature \cite{Zhou2018}. 
The architecture of the GNN used in this work resembles the ParticleNet model proposed by Qu et al. \cite{Qu2019}. It consists out of three Edge Convolutional blocks \cite{wangDynamicGraphCNN2019}, followed by a global pooling layer and two fully connected layers. A custom open-source implementation for this architecture was developed using the tensorflow \cite{Abadi2016} back end\footnote{See \url{https://github.com/StefReck/MEdgeConv}.}.

\section{Atmospheric muon reconstruction}
\label{sec:muons}
\begin{figure}
  \centering
  \begin{subfigure}[b]{0.49\linewidth}
    \includegraphics[page=3,width=\linewidth]{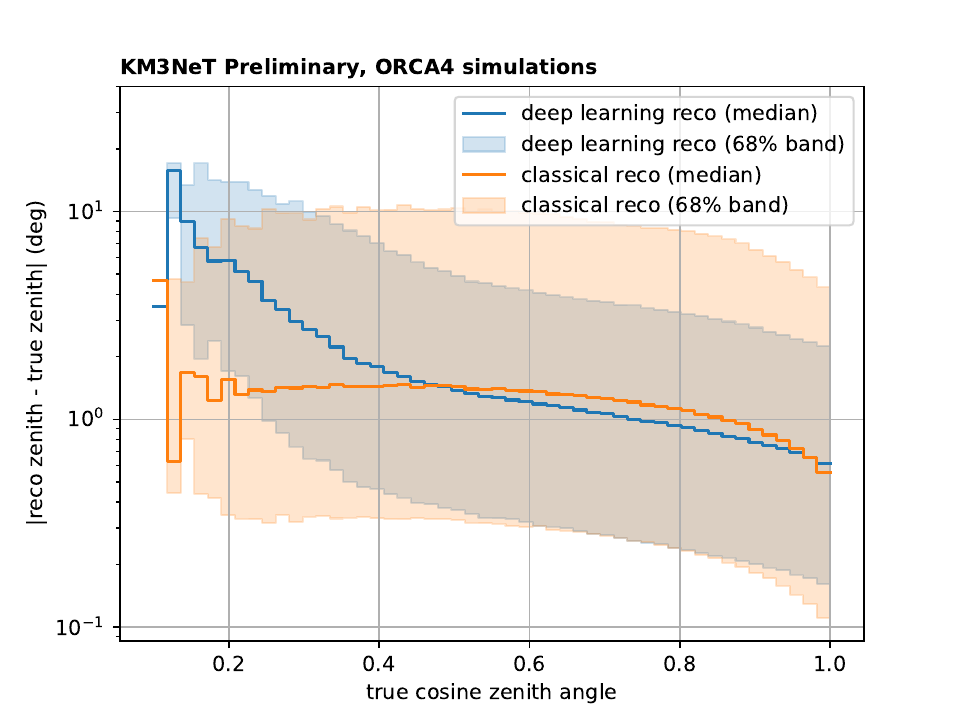}
    \caption{all events}
  \end{subfigure}
  \begin{subfigure}[b]{0.49\linewidth}
    \includegraphics[page=4,width=\linewidth]{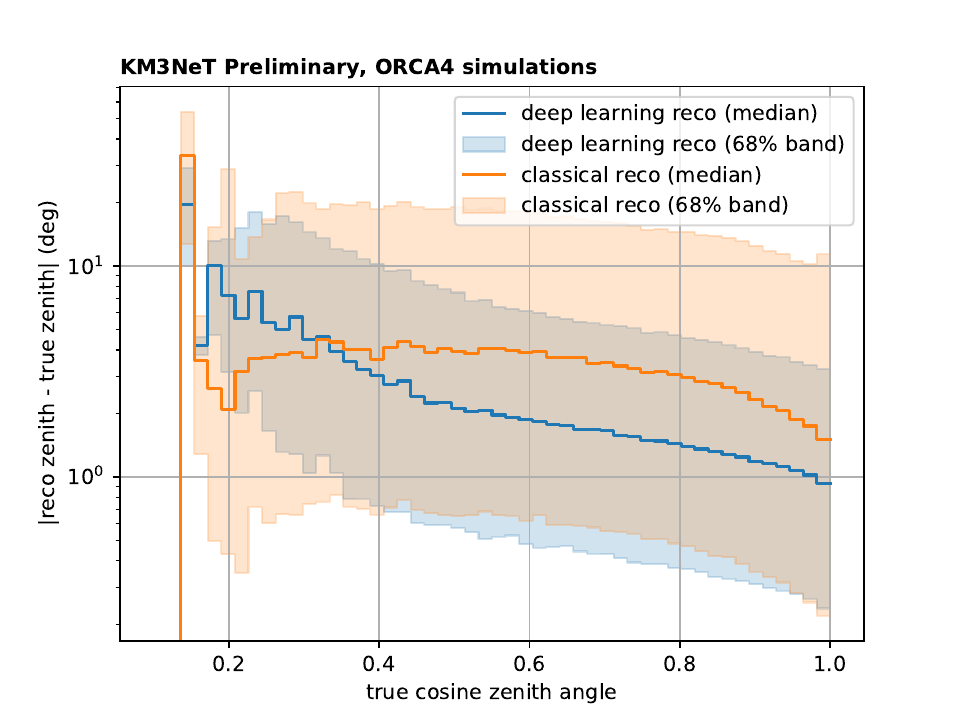}
    \caption{events with two or more muons}
  \end{subfigure}
  \caption{
    Absolute difference between reconstructed and true zenith angle plotted over the true zenith angle for
    selected atmospheric muons in ORCA4.
	Shown are the median and the 68\% band for the classical reco (orange) and the deep learning reco
	(blue). 
  }
  \label{fig:zen_mae}
\end{figure}

KM3NeT’s primary goal is to study neutrinos. However, the majority of detected events are atmospheric muons produced in cosmic ray induced air showers, which usually pose as a background for neutrino measurements. But they provide a wide range of applications as well, like
measuring the cosmic ray composition.
GNNs can be used to reconstruct the zenith angle and the muon multiplicity of an incident muon or muon bundle. 
For this, a set of 25 million atmospheric muon events was simulated using the software package MUPAGE \cite{Carminati2008} for the initial four-line set-up of the ORCA detector.
\autoref{fig:zen_mae} shows the zenith angle reconstruction performance in comparison to a classical maximum-likelihood based algorithm for single tracks as described in \citep{Adrian-Martinez2016}. 
Since it was trained on the expected distribution, the deep learning reconstruction is biased for true cosine zenith below 0.5, leading to an increase in the error there. GNNs provide a comparable precision to classical methods on atmospheric muon events and a substantial reduction of the median zenith error by a factor of 2 for the rarer multi-muon events, which make up about a fifth of the dataset.
Since the classical method was not optimised for multi-muon events, its performance could potentially be improved, e.g. by an adjustment of the used PDF.

\begin{figure}
\centering
\begin{minipage}[t]{.45\textwidth}
  \centering
  \captionsetup{width=.95\linewidth}
  \includegraphics[page=6,width=\linewidth]{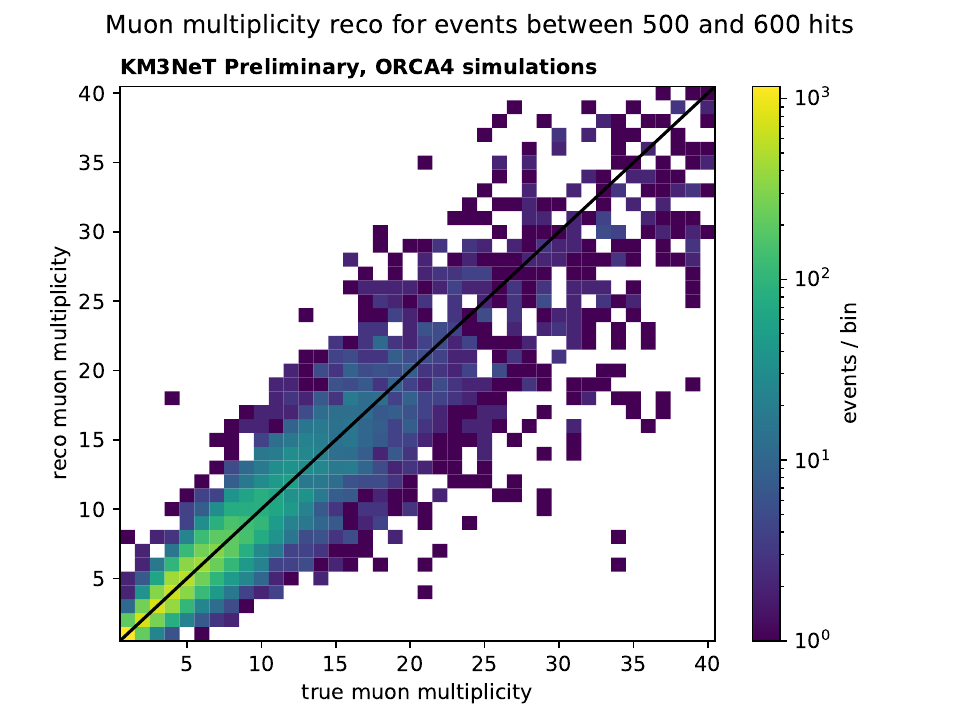}
  \captionof{figure}{
Reconstructed versus true muon multiplicity for events with a similar number of hits (500 to 600). 
Even without use of the correlation between the multiplicity and the number of hits, the reconstruction is meaningful.
	}
  \label{fig:muon_2d}
\end{minipage}%
\hfill
\begin{minipage}[t]{.45\textwidth}
  \centering
  \captionsetup{width=.95\linewidth}
  \includegraphics[page=7,width=\linewidth]{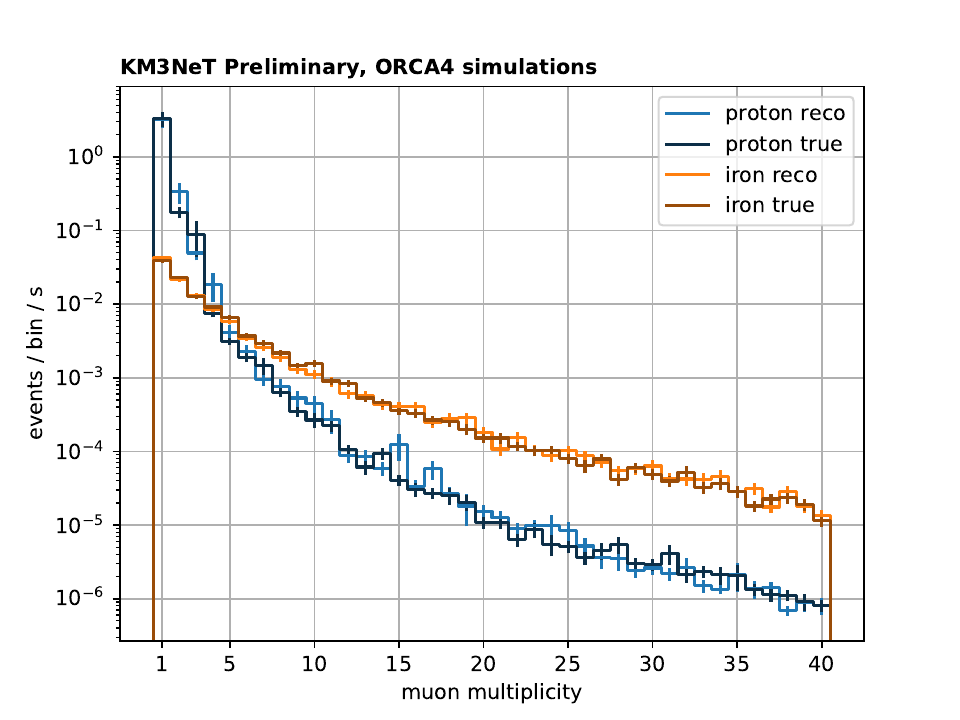}
  \captionof{figure}{
Reconstructed (light) and true (dark) muon multiplicity rates
for events generated in proton (blue) and iron (orange)
induced atmospheric showers.   
  }
  \label{fig:muon_corsika}
\end{minipage}
\end{figure}

The number of atmospheric muons in an event (muon multiplicity) can be used for the identification of primary particles.
This deep learning approach is the first reconstruction of the muon multiplicity in KM3NeT. 
Even though the detector is still in an early stage of construction, deep learning can already provide a good estimator for the muon multiplicity (see \autoref{fig:muon_2d}).
 \autoref{fig:muon_corsika} shows the result of applying the trained network on 250,000 atmospheric muon events simulated with Corsika \cite{Heck:1998vt} SIBYLL 2.3c \cite{Riehn2017} using the GST-3 \cite{Gaisser2013} spectrum. 
As can be seen, the current detector already provides a decent separation power between iron and proton induced events.  

\section{Studies on neutrino selections}
\label{sec:neutrino}
GNNs can be used to select neutrino events in ORCA4 real data. To this end, different networks have been trained on simulated neutrino data \cite{gseagen} to perform classification and reconstruction tasks. The cuts are optimized on MC data for the highest possible neutrino yield while keeping a low background contamination. The application to real data shows a reasonable agreement between expectation and observation, proving to be a valuable method to cross-check conventional methods.

\begin{figure}
\centering
\begin{minipage}[t]{.45\textwidth}
  \centering
  \captionsetup{width=.95\linewidth}
    \includegraphics[width=0.99\linewidth]{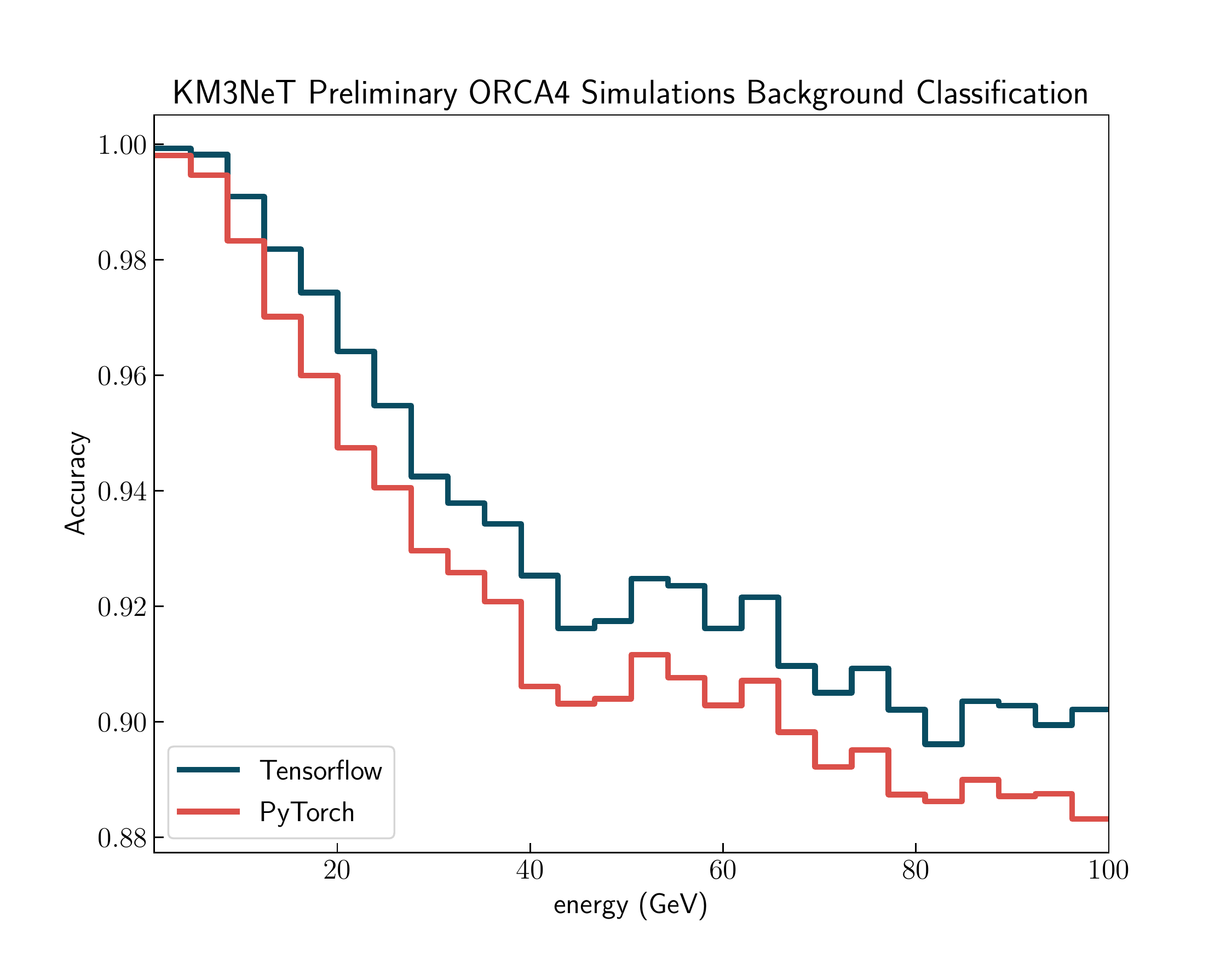}
  \captionof{figure}{A comparison of the accuracy for background muon classification between PyTorch and Tensorflow for the same ParticleNet architecture \cite{Qu2019}.}
  \label{fig:pytorch}
\end{minipage}%
\hfill
\begin{minipage}[t]{.45\textwidth}
  \centering
  \captionsetup{width=.95\linewidth}
  \includegraphics[width=1.0\textwidth]{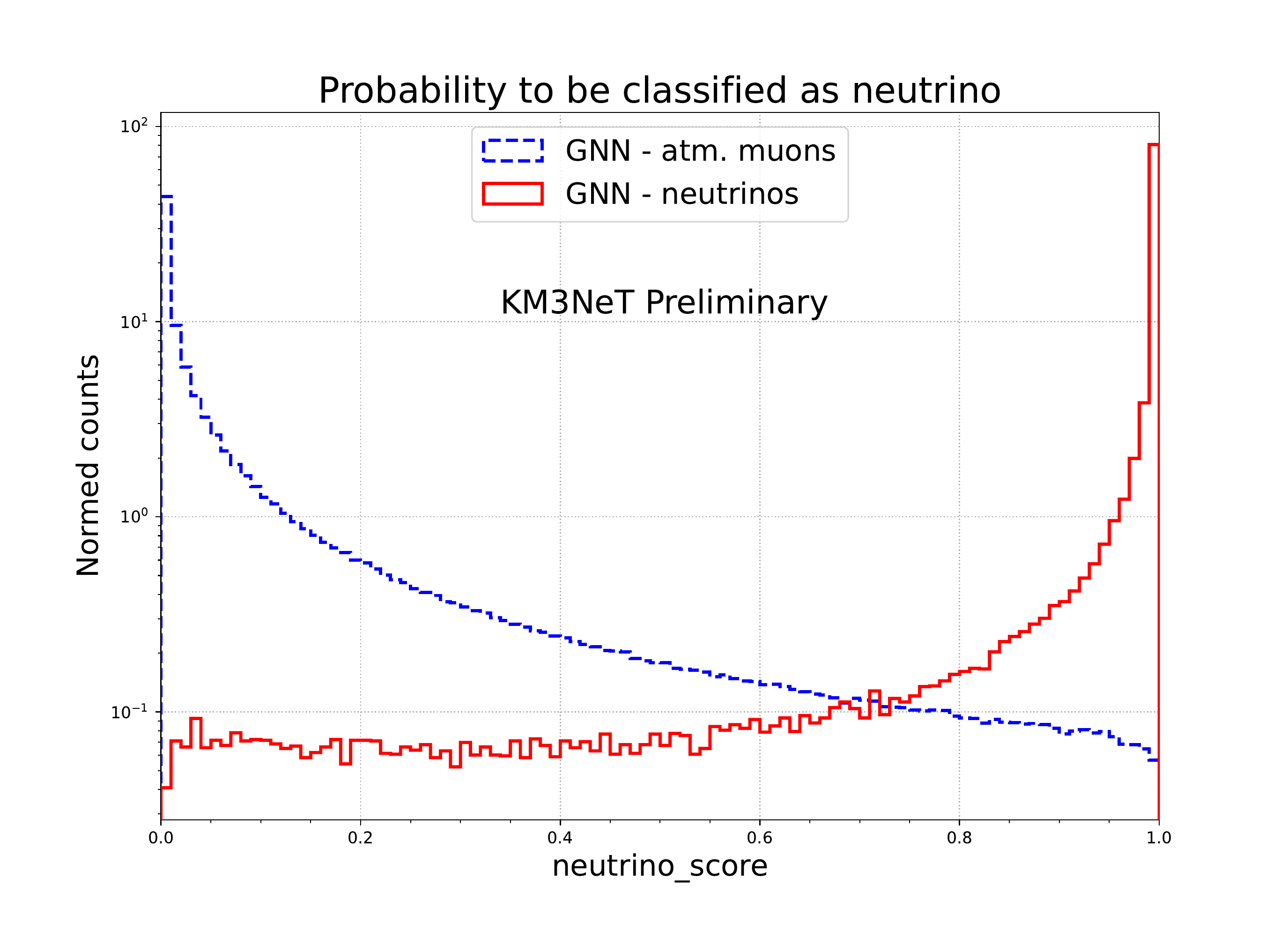}
  \captionof{figure}{Normed distribution of the \texttt{neutrino\_score} for true muon (blue-dashed) and true neutrino (red-continuous) events for the full ARCA detector.  
  }
  \label{fig:score_nu_mu}
\end{minipage}
\end{figure}

The ParticleNet used this far is implemented in Tensorflow. An alternative architecture using PyTorch Geometric \cite{Fey/Lenssen/2019} has also been tested. 
The benefit of this framework is that it has many implementations of GNN layers included,
which allow for easy prototyping of different architectures. As a first benchmark for the framework, ParticleNet is implemented and it is trained for the signal/background classifier on ORCA4.
The accuracy of both implementations 
can be seen in \autoref{fig:pytorch}. The 
difference between the implementations
can be attributed to the use of more advanced learning rate schedulers in Tensorflow.
The shape of the accuracy curve is identical between the two implementations.
GNNs can also be used for selecting cosmic neutrinos in ARCA over the muon background. This can be achieved by cutting over a GNN \texttt{neutrino\_score}, representing the probability of an event being induced by a neutrino. The selected neutrinos are then further classified into tracks and showers by a subsequent GNN classification step. \autoref{fig:score_nu_mu} shows the distribution of the \texttt{neutrino\_score} assigned by the GNN to true neutrinos and true atmospheric muons, for the full ARCA detector. The GNN performances are now being compared with other ML methods with the goal of finding the most promising method for cosmic neutrino selection.

\bibliographystyle{JHEP}
\bibliography{bibliography}

\end{document}